\documentclass{ws-ijmpcs}
\usepackage[bookmarks=true,colorlinks=true,citecolor=blue]{hyperref}
\newcommand{\be}{\begin{equation}}
\newcommand{\en}{\end{equation}}
\newcommand{\bea}{\begin{eqnarray}}
\newcommand{\ena}{\end{eqnarray}}
\newcommand{\lbl}[1]{\label{eq:#1}}
\newcommand{\lbltab}[1]{\label{tab:#1}}
\newcommand{\lblfig}[1]{\label{fig:#1}}

\newcommand{\rf}[1]{(\ref{eq:#1})}
\newcommand{\Table}[1]{\ref{tab:#1}}
\newcommand{\fig}[1]{\ref{fig:#1}}

\newcommand{\bc}{\begin{center}}
\newcommand{\ec}{\end{center}}
\newcommand{\bt}{\begin{tabular}}
\newcommand{\et}{\end{tabular}}
\newcommand{\bg}{\begin{minipage}}
\newcommand{\eg}{\end{minipage}}
\newcommand{\ba}{\begin{array}}
\newcommand{\ea}{\end{array}}
\newcommand{\qdd}{q^2}
\newcommand{\om}{{Omn\`es}\ }
\newcommand{\mpid}{m_\pi^2}
\newcommand{\mpi}{m_\pi}
\newcommand{\pip}{\pi^+}
\newcommand{\pim}{\pi^-}
\newcommand{\piz}{\pi^0}

\newcommand{\re}{{\rm Re\,}}
\newcommand{\Kbar}{\bar{K}}
\newcommand{\mvd}{M_V^2}
\newcommand{\mv}{M_V}
\newcommand{\gv}{\Gamma_V}
\newcommand{\betapi}{\sigma_\pi}
\newcommand{\Jhat}{ \hat{J}^{I,\pi} }

\newcommand{\lapprox}{%
\mathrel{%
\setbox0=\hbox{$<$}\raise0.6ex\copy0\kern-\wd0\lower0.65ex\hbox{$\sim$}}}
\newcommand\TT{\rule{0pt}{2.0ex}}         
\newcommand\BB{\rule[-1.0ex]{0pt}{0pt}}  

\begin{document}

\markboth{B. Moussallam}{Unified dispersive approach to
  $\gamma^*\to\gamma\pi\pi$ and $\gamma\gamma\to\pi\pi$}

%
\catchline{}{}{}{}{}
%

\title{UNIFIED DISPERSIVE APPROACH TO $\gamma^*\to\gamma\pi\pi$
AND $\gamma\gamma\to\pi\pi$ }

\author{B. Moussallam}

\address{Groupe de physique th\'eorique, IPN b\^at. 100,  Universit\'e Paris-Sud
  11, Orsay Cedex, France}


\maketitle

\begin{history}
\received{Day Month Year}
\revised{Day Month Year}
\end{history}

\begin{abstract}
A  representation of the amplitude $\gamma^*(\qdd)\to\gamma\pi\pi$
is proposed  which combines large $N_c$ chiral resonance Lagrangian
modelling with general unitarity and analyticity properties. 
The amplitude is constrained from $\gamma\gamma$ scattering 
results and $e^+ e^- \to\gamma\pi^0\pi^0$ measurements by the CMD-2 and SND
collaborations. As an application, the contribution of the
$\pi\pi+\gamma$ states in the HVP contribution to the muon $g-2$ are
reconsidered, taking into account the effect of the strong $S$-wave
$\pi\pi$ rescattering in a model independent way.  

\end{abstract}


\section{Introduction}	
The leading hadronic contribution to the muon $g-2$ is associated with
the hadronic vacuum polarization (HVP) function, and the contribution from
$\pi^+\pi^-$, proportional to the square of the pion form factor,
dominates the  HVP unitarity relation.
This has triggered  experimental efforts for measuring  
the pion form factor to high accuracy, in particular, via
the initial-state radiation (ISR)
method (see~\cite{Babusci:2012rp,Lees:2012cj} and references therein).  
In the  $e^+ e^- \to \gamma\pi^+ \pi^- $ cross-section, the final-state
radiation (FSR) amplitude contributes in addition to the ISR. In
principle, they could be separated experimentally by performing a
partial-wave analysis. 
The FSR amplitude is also needed for computing the $\gamma\pi\pi$
contribution in the HVP unitarity relation.
In practice, the FSR amplitude is often estimated using the   sQED
approximation, which treats the pions as point-like and non-interacting. 
It ignores, in particular, the influence of the strong $\pi\pi$
$S$-wave attraction at low energy. A modelling of this effect using a narrow
$\sigma$-meson gives  surprisingly large results~\cite{Narison:2003ur}. We
discuss here an approach in which $\pi\pi$ rescattering is treated in the model
independent \om method~\cite{Omnes:1958hv}. It can be viewed as a
generalization of  classic work on the
$\gamma\gamma\to\pi\pi$ 
amplitude~\cite{Gourdin:1960,Morgan:1987wi,Donoghue:1993kw} 
and uses $\gamma\gamma$ scattering experimental results as constraints.
A further generalization to the case of two virtual photons is
presented at this conference~\cite{Hoferichter}, which will be applied
to the  light-by-light hadronic contribution to the $g-2$.    

\section{Analyticity of partial-waves when $\qdd\ne0$ } 
The \om method applies to partial-wave projected amplitudes, it 
combines the  unitarity relation and  analyticity properties. 
We restrict ourselves here to the elastic scattering region $s\lapprox
1$ GeV$^2$ which will limit the applicability of the amplitude to
virtualities $\qdd\lapprox 1$ GeV$^2$.
In the case of two real
photons, the partial-wave amplitude is an analytic function of the
$\pi\pi$ energy $s$,  except for two cuts,
the right-hand cut which lies on $[4\mpid,\infty]$ and the left-hand cut on
$[-\infty,0]$. The discontinuity across the right-hand cut is given by
the unitarity relations, these have exactly the same form for $\gamma\gamma$
and $\gamma^*\gamma$ amplitudes. In contrast, the left-hand cuts 
differ. The main issue is to properly define this cut and verify that
no anomalous threshold is present. 

The left-hand cut is associated with singularities of the unprojected
amplitude in the crossed channels. One firstly has the pion pole in the
$\gamma^* \pi^+\to \gamma\pi^+$ amplitude (so-called Born amplitude)
the $J=0$ projection reads,
\begin{equation}\lbl{bornamplit}
{h}^{Born}_{0,++}(s,\qdd)={F_\pi^v(\qdd)\over s-\qdd}\,\Big[
\dfrac{4\mpid}{\betapi(s)} L_\pi(s) -2\qdd \Big],\ \
L_\pi(s)=\log{ 1 + \betapi(s)\over1 -\betapi(s)}
\end{equation}
with $\betapi(s)=\sqrt{1-4\mpid/s}$. 
Having  $\qdd\ne0$ affects the amplitude through the pion
form factor $F_\pi^v(\qdd)$ but also  the singularities: the Born amplitude
displays a pole at $s=\qdd$ in addition to a left-hand cut. Using
the $\qdd+i\epsilon$ prescription moves the pole
away from the right-hand cut when $\qdd > 4\mpid$.
Secondly, one must consider the cuts associated with $\gamma^*\pi\to
n\pi \to  \gamma\pi$.  These processes are expected
to display sharp resonance effects below 1 GeV from the
vector mesons $\rho$, $\omega$. We may start with a large $N_c$
approximation, where resonances generate simple poles in $\gamma^*\pi
\to  \gamma\pi$ (note that scalar mesons, which violate large $N_c$
rules are not allowed). Using a resonance chiral Lagrangian, the
contributions from a vector meson exchange to the three independent
invariant amplitudes read
\begin{equation}\lbl{vectoramplit}
\ba{l}
 A^V(s,t,\qdd)={\tilde{C}_V}{ F_{V\pi}(\qdd)}\!\Big[
 \dfrac{s-4m_\pi^2-4t+\qdd}{ t-\mvd}
+\dfrac{s-4m_\pi^2-4u+\qdd}{ u-\mvd}\!\Big]   \\
 B^V(s,t,\qdd)={\tilde{C}_V}{ F_{V\pi}(\qdd)}\!\Big[
\dfrac{1}{2(t-\mvd)} +\dfrac{1}{2(u-\mvd)}\Big] \\
 C^V(s,t,\qdd)={\tilde{C}_V}{ F_{V\pi}(\qdd)}\!\Big[
\dfrac{1}{ t-\mvd} -\dfrac{1}{ u-\mvd}\Big]\\
\ea
\end{equation}
The main difference when $\qdd\ne0$ is from the kinematics: for
$\gamma\gamma\to\pi\pi$ the Mandelstam variables $t$, $u$ are negative
while for $\gamma^*(\qdd)\to \gamma\pi\pi$ they
lie in the range: $[4\mpid,(\sqrt{\qdd}-\mpi)^2]$. 
One must then take the width of the resonance into account, and this
must be done in a way  consistent with the general analyticity properties
for, otherwise, the \om method would not be applicable. This may be
implemented by using a K\"all\'en-Lehmann representation, i.e. by
replacing, in eq.~\rf{vectoramplit}  
\be\lbl{kallenlehman}
{1\over\mvd-w}\longrightarrow 
\widetilde{BW}_{\!\!V}(w)={1\over\pi}\int_{4\mpid}^\infty dt'
{\sigma(t',\mv,\gv)\over t'-w},\quad w=t,u
\en
The function $\widetilde{BW}_{\!\!V}$ has a cut on the
first Riemann sheet, while a pole  appears on the second sheet. 
The cut structure of the partial-wave projection of the vector-exchange
amplitude is illustrated on figs.~\fig{complexcut}: the left figure 
shows that the cut extends into the complex plane and approaches
the right-hand cut. The vicinity of
the right-hand cut is illustrated on the right figure. Thanks to the analytic
propagator and the $\qdd+i\epsilon$ prescription, no intersection
actually occurs, which guarantees the absence of an anomalous threshold
and the applicability of the usual \om method.  \\[-0.5cm] 
\begin{figure}[pt]
\centerline{\includegraphics[width=0.45\linewidth]{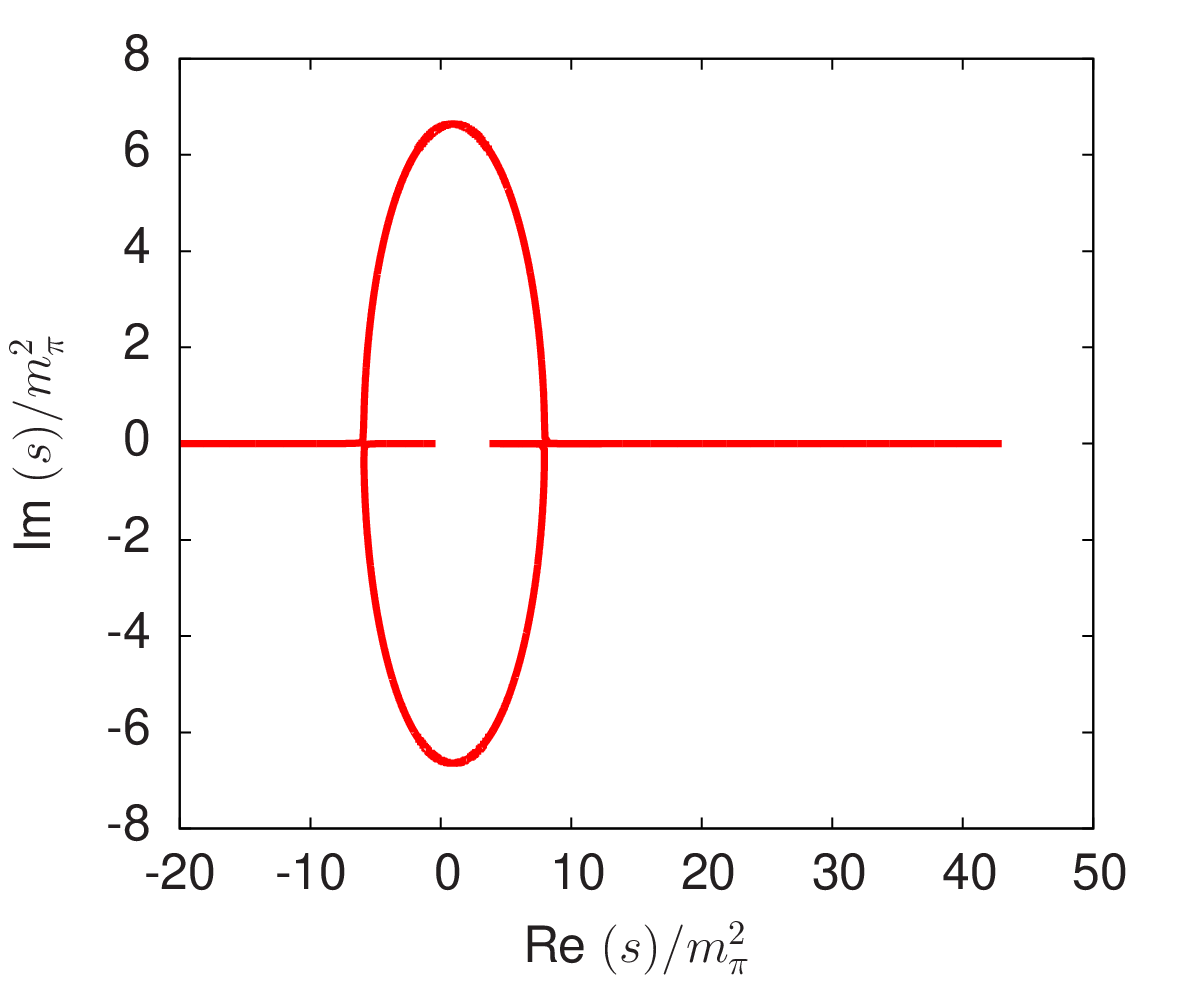}\includegraphics[width=0.45\linewidth]{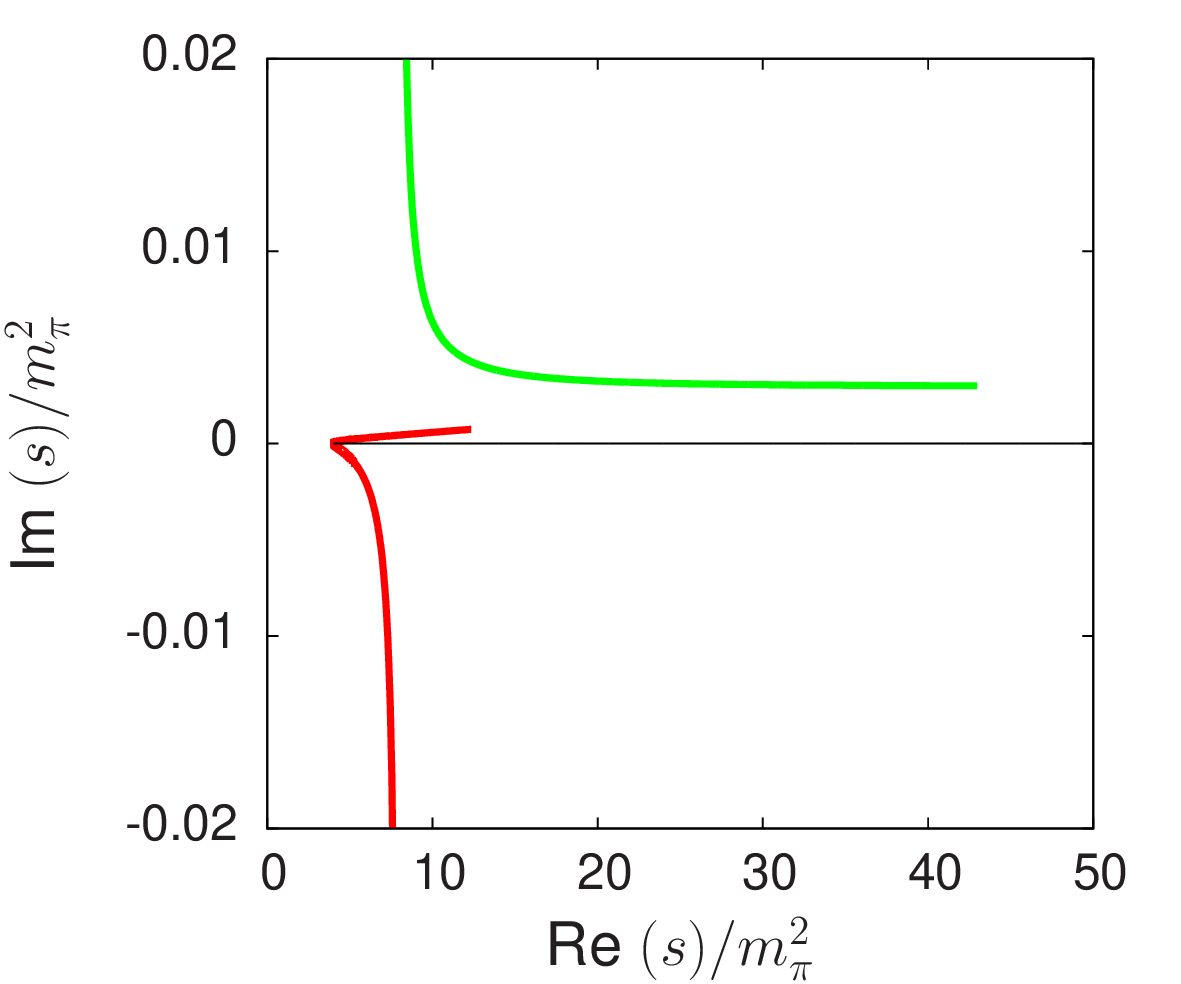}}
\caption{Cut structure of a partial-wave 
 projection of the   vector-exchange amplitude~\rf{vectoramplit}.}
\lblfig{complexcut}
\end{figure}
\section{Soft photon, chiral and experimental constraints}
We consider an \om representation for the
$\gamma^*\to\gamma(\pi\pi)_I$ amplitudes based on twice-subtracted
dispersion relations i.e. involving two polynomial parameters for each
isospin $I$. These  account e.g. for higher mass resonances
not explicitly included and depend on $\qdd$. 
Beside the properties of analyticity and elastic unitarity, there are
additional physical constraints that must be imposed. 
Gauge invariance imposes that the amplitudes minus the Born term 
must vanish in the soft photon limit~\cite{Low:1958sn}
which eliminates one of the parameters for each $I$. 
The helicity amplitude $H^I_{++}(s,\qdd,z)$, where $z$ is the cosine
of the scattering angle in the $\pi\pi$ CMS, can be written as
\be\lbl{omnesamplit}
H^I_{++}(s,\qdd,z)= 
{H}^{I,Born}_{++}(s,\qdd,z)
+{\sum_{V=\rho,\omega}}\,{H}^{I,V}_{++}(s,\qdd,z) +H^{I,resc}_{++}(s,\qdd,z)
\en
The  last term in eq.~\rf{omnesamplit} accounts for the rescattering in
the $J=0$ partial-wave, it reads
\be\ba{l}
H^{I,resc}_{++}(s,\qdd,z)= \Omega_0^I(s) \Big\{ (s-\qdd)\, b^I(\qdd)
+ s\,{F^v_\pi(\qdd)} \Big[ 
\dfrac{s\,(J^{I,\pi}(s,\qdd)-J^{I,\pi}(\qdd,\qdd))}{ s-\qdd } \\
\phantom{H^{I,resc}_{++}()}
-\qdd \Jhat(\qdd) \Big]        
+{s} {\displaystyle\sum_{V=\rho,\omega}} {F_{V\pi}}(\qdd)
\Big[ s\, J^{I,V}(s,\qdd)-\qdd  J^{I,V}(\qdd,\qdd)\Big]\Big\}\ .\\
\ea\en
$\Omega_0^I$ is the  usual \om function and
$J^{I,\pi}$, $J^{I,V}$ are the related integrals (see~\cite{Omnes:1958hv})
involving  the partial-wave projections
of the Born and the vector-exchange amplitudes respectively. Finally,  
$\Jhat=\partial{J}^{I,\pi}(s,\qdd)/\partial{s}$ at $s=\qdd$.  The
other helicity amplitudes $H_{+-}$, $H_{+0}$ are
affected by rescattering from $J\ge 2$  partial-waves.
\begin{figure}[pt]
\includegraphics[width=\linewidth]{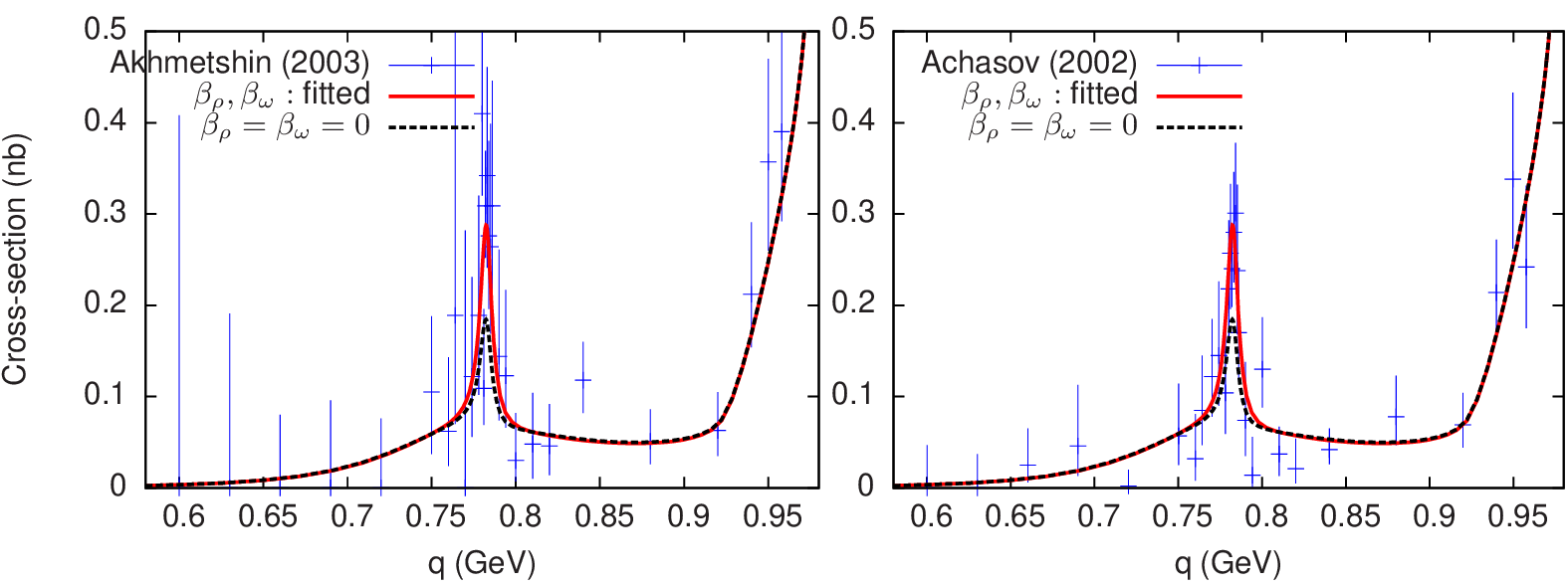}
\caption{Experimental results for $\sigma(e^+ e^- \to
  \gamma\pi^0\pi^0)$ and our two parameter fit.}
\lblfig{resultsigpi0pi0}
\end{figure}

The two functions $b^0(\qdd)$, $b^2(\qdd)$,
are constrained by chiral symmetry. In the
exact chiral limit, the $\gamma^*\gamma$ amplitude for producing a
$\pi^0$ pair satisfies a soft pion theorem: it vanishes at
$s=0$ for any value of $\qdd$.  In the real world, the Adler zero
moves to $s_A=O(\mpid)$ and depends on $\qdd$. The correct chiral
behaviour is enforced by matching the dispersive amplitudes for both
$\pi^0\pi^0$ and $\pi^+\pi^-$ with the corresponding chiral expansion
expressions~\cite{Donoghue:1988eea} at $s=0$ and small
$\qdd$.  For larger values of $\qdd\lapprox 1$ GeV$^2$ the dependence
is dominated  by the light vector resonances. Introducing the
combinations $b^n=(-b^0+\sqrt2 b^2)/\sqrt3$ and $b^c=-(\sqrt2
b^0+b^2)/\sqrt6$ corresponding to the $\pi^0\pi^0$ and $\pi^+\pi^-$
channels, the following parametrization encodes these properties
\be\lbl{parambnbc}
b^n(\qdd)= b^n(0)F_\chi(\qdd) + F_R(\qdd),\quad
b^c(\qdd)= b^c(0)            + F_R(\qdd)
\en
where  $F_R(\qdd)=\beta_\rho (GS_\rho(\qdd)-1)+\beta_\omega
(BW_\omega(\qdd)-1)$ 
involves the Gounaris-Sakurai and Breit-Wigner functions, and
$F_\chi(\qdd)=12\mpid[\mpid L^2_\pi(\qdd)/\qdd +\betapi(\qdd)L_\pi(\qdd)+3]$. 
The values of $b^n(0)$, $b^c(0)$ are determined from the
polarizabilities of the $\pi^+$ and the $\pi^0$ which we take to be
compatible with the chiral predictions. In the
parametrization~\rf{parambnbc}  we used the same resonance function 
for $b^n$ and $b^c$ i.e. we neglected the resonance contribution to
$b^2$. This is justified from the fact that
the \om functions satisfy the inequality $\vert\Omega_2(s)\vert <<
\vert\Omega_0(s)\vert$  in the physically relevant region 
$4\mpid\le s\le \qdd$ which suppresses the influence of $b^2$. 
Thanks to this simplification, determining the two parameters $\beta_\rho$,
$\beta_\omega$ from $\sigma(e^+ e^-\to \gamma\pi^0\pi^0)$ allows one
to predict $\sigma^{FSR}(e^+ e^-\to \gamma\pi^+\pi^-) $. 
A combined fit to the two data sets from the SND and CMD-2 
collaborations~\cite{Achasov:2002jv,Akhmetshin:2003rg}
gives: $\beta_\rho= 0.05\pm 0.09$, 
$\beta_\omega=(-0.37\pm0.09)\cdot10^{-1}$ GeV$^{-2}$ with
$\chi^2/N_{dof}=38/50$, this is illustrated in Fig.\fig{resultsigpi0pi0}.

\vspace*{-0.3cm}
\section{Application to the  $\pi\pi\gamma$ contributions to the muon $g-2$}
The contribution of the HVP to the muon $g-2$ which involve two pions
plus one photon can be written in terms of infrared finite cross-sections,
\be\lbl{g-2integral}
\left.{g-2\over2}\right\vert_{\pi\pi\gamma}
={1\over4\pi^2}\int_{4\mpid}^\infty d\qdd K_\mu(\qdd)\Big(
\sigma^{sQED}_{e^+ e^-\to\pi^+\pi^-  \gamma}(\qdd) + \sum_{n,c}
\hat{\sigma}^{n,c}_{e^+ e^-\to{\pi\pi} \gamma}(\qdd) \Big)
\en
where (see e.g.~\cite{Jegerlehner:2009ry} for the explicit expression
of the functions $K_\mu$ and $\eta$)
\be\ba{l}
\sigma^{sQED}_{e^+ e^-\to\pi^+\pi^- \gamma}= 
\dfrac{\alpha^3}{3\qdd}\sigma^3_\pi(\qdd)\vert F_\pi^v(\qdd)\vert^2
\times\eta(\qdd) \\[0.2cm]
\hat{\sigma}^{c,n}_{e^+ e^-\to{\pi\pi} \gamma}
=\dfrac{\alpha^3}{12(\qdd)^3}{\displaystyle\int_{4\mpid}^{\qdd}}
ds(\qdd-s)\sigma_\pi(s){\displaystyle\int_{-1}^1 }dz 
( \vert\hat{H}^{c,n}_{++}\vert^2
+ \vert\hat{H}^{c,n}_{+-}\vert^2 
+ \vert\hat{H}^{c,n}_{+0}\vert^2 )
\ea\en
where $\hat{H}^n_{\lambda\lambda'}= H^{n,V+resc.}_ {\lambda\lambda'}$
and $\vert \hat{H}^c_{\lambda\lambda'}\vert^2= 2\re(
H^{*Born}_{\lambda\lambda'} H^{c,V+resc.}_{\lambda\lambda'})+ \vert
H^{c,V+resc.}_{\lambda\lambda'}\vert^2$. 
\begin{table}[ph]
\tbl{$\pi\pi\gamma$ contributions (central values) to the muon $g-2$ from the
  integration region $\sqrt{\qdd} \le 0.95$ GeV.}
{\bt{ll|c}\hline
\TT\BB channel & cross-section  & $(g-2)/2$\\ \hline
\TT {$\gamma\pip\pim$} & $\vert H^{Born}\vert^2$ &  $41.9\times 10^{-11}$ \\ 
 {$\gamma\pip\pim$} & ${H^*}^{Born} H^{V+resc}$ & $(1.31\pm0.30) \times 10^{-11}$\\
 {$\gamma\pip\pim$} & $\vert H^{V+resc}\vert^2$ & $(0.16\pm0.05) \times
10^{-11}$ \\  
 {$\gamma\piz\piz$} & $\vert H^{V+resc}\vert^2$& $(0.33\pm0.05) \times 10^{-11}$ \\ \hline
\et\lbltab{g-2contribs}}
\end{table}

The contributions to the muon $g-2$, 
restricting the  integration range in eq.~\rf{g-2integral} 
to $\sqrt{\qdd} \le 0.95$ GeV, within the domain of validity of the model, 
are shown in table~\Table{g-2contribs}. As compared to previous work,
we find for the contribution linear in $H^{Born}$ a different sign
than ref.~\cite{Dubinsky:2004xv}, which is due to the effect of the
rescattering. The results from the last two lines can be
compared with ref.~\cite{Narison:2003ur} who use a sigma resonance
approximation: our result is smaller in magnitude by a factor of three.
The table shows that the sQED contribution is largely dominant. Still,
it would be of interest to be able to extend the integration range somewhat   
since one expects a kinematical increase of $\sigma^{FSR}$ when $\sqrt{\qdd}
> m_\omega+m_\pi$. This would necessitate to include $J=2$
rescattering, which is easy to implement, and  also account for
$\pi\pi-K\Kbar$ inelasticity. 

\vspace*{0.3cm}
\noindent{\bf Acknowledgements:\ }{\small Supported by the European
Community-Research Infrastructure Integrating Activity "Study of
Strongly Integrating Matter" (acronym HadronPhysics3)}
\vspace*{-0.4cm}
\bibliography{essai}
\bibliographystyle{ws-ijmpcs}
\end{document}